# Sustainability, behavior patterns and crises

R. Vilela Mendes [1]

CEMS.UL, Faculdade de Ciências, Universidade de Lisboa
Academia das Ciências de Lisboa

## Abstract

Sustainability has been defined as *meeting the needs of the present without compromising the ability of future generations to meet their own needs*. But what are the needs of the present? And are they met? From the poor performance of the 2030 Sustainable Development Goals (SDG), defined by the UN in 2015, not even the collective needs of the present seem to be met. How to expect not to compromise the needs of the future? Is the achievement of global world goals incompatible with the characteristic processes of human evolution, as some authors have recently suggested?

Simple mathematical models cannot capture the whole breadth of human experience and destiny. But, on the other hand, one should not neglect whatever insights they may provide. And what these models teach us is how the behavior pattern "Parochial cooperation - Conflict - Growth" was reached and how this pattern, in addition to leading to several types of crises, is also on the way of the global governance needed to achieve the SDG's

## Table of contents

---

[1] rvilela.mendes@gmail.com; rvmendes@ciencias.ulisboa.pt;
https://label2.tecnico.ulisboa.pt/vilela/

## I. Sustainability: The goals

What is sustainability? The Brundtland report, of the World Commission on Environment and Development, proposed in October 1987 the following definition:

> *"Meeting the needs of the present without compromising the ability of future generations to meet their own needs"*

In 2015 the United Nations announced 17 Sustainable Development Goals, to be achieved by 2030, which adequately define the *"needs of the present"*. One may ask: Are they being achieved? Is there any real progress since 2015? Not really! For example:

**Goal no.2: Zero hunger,** hunger and malnutrition being defined as *"a situation that exists when people lack secure access to sufficient amounts of safe and nutritious food for normal growth and development and an active and healthy life.* (UN Food and Agriculture Organization, 2010)
\# In 2015, when the 17 Sustainable Development Goals were defined, the percentage of undernourished population was 7.9% (584.6 million people)
\# In 2022 it was 9.2% (731.4 million people)
(Data source: FAO, UN - 2023)

**Goal no. 10: Reduced inequalities**

The Gini coefficient is $A/(A+B)$, $A$ and $B$ being the areas above and below the Lorentz curve. The Lorentz curve is the "cumulative share of income earned" as a function of the "cumulative share of people from lowest to highest incomes". Between 2015 and 2022, the variation of the Ginni index differs from region to region but the world average remained the same = 0.67.
(Data source: World Inequality Database (WID.world), 2024)

**Goal no.13: Climate action** - This has a central role and impact on all SDG's
 * Mean atmospheric $CO_2$ at the Mauna Loa Observatory: In 2015: 402 ppm; in 2024: 422 ppm
(Data source: Scripps Institution of Oceanography, NOA Global Monitoring Laboratory)
 * Annual $CO_2$ emissions (fossil fuels and industry): In 2015: 35.7 Gt; in 2023: 37.5 billion tonnes
(Data source: Global Carbon Budget, 2023)

**Goal no. 16: Peace, justice and strong institutions**

\# **Armed conflicts** (UCDP-Uppsala and PRIO-Oslo)
*A disagreement between organized groups, or between one organized group and civilians, that causes at least 25 deaths during a year. This includes combatant and civilian deaths due to fighting, but excludes deaths due to disease and starvation resulting from the conflict.*
In 2015: 103608 deaths; in 2022: 204009 deaths (UCDP)

# Justice and strong institutions

Representation vs. rule of law:

|   | Participation (2015) | (2022) | Rule of law (2015) | (2022) |
|---|---|---|---|---|
| *Russia* | 0.36 | 0.31 | 0.35 | 0.33 |
| *China* | 0 | 0 | 0.38 | 0.35 |
| *SaudiArabia* | 0 | 0 | 0.39 | 0.35 |
| *USA* | 0.77 | 0.67 | 0.78 | 0.7 |
| *Canada* | 0.8 | 0.82 | 0.83 | 0.78 |
| *Germany* | 0.92 | 0.87 | 0.93 | 0.91 |
| *Portugal* | 0.89 | 0.82 | 0.75 | 0.66 |

(Data from International IDEA, Global State of Democracy Indices)

From 2015 to 2022 the situation seems to have become worse, except, of course, in China and Saudi Arabia where the degree of participation remained stable (at zero).

etc.

**In conclusion:** If not even the *"needs of the present"* are met, how could one expect any consideration for the needs of the future? However, this is a very serious matter because, meeting the needs of the present, is perhaps the best way not to compromise the needs of the future.

The 2030 SDG's of the UN are a nice inspired invention of the human mind. Why are they not being implemented? Two hypothesis:

1 - They are irrelevant nonsensical do-gooder wishes, incompatible with human nature.

2 - There is nobody to enforce them.

Both hypothesis are probably true.

1 - Is fulfilling the goals really incompatible with the human evolution traits? Some authors think so [1] [2]. In the next section using simple replicator dynamics models, this question will be analyzed. One finds how humans, evolving in interaction with other humans, have developed a behavior pattern that shapes the vast majority of their social interventions as well as their interactions with Nature.

2 - The question of whether it is possible or impossible to enforce the SDG's, will be dealt with in the final section.

## II. P_Cooperation-Conflict-Growth: Evolutionary emergence of a behavior pattern

## II A. The fragility of cooperation

Humans are a highly cooperative species [3]. Cooperative in helping each other, cooperative in achieving material and intellectual goals unmatched by other species, also cooperative in war and genocide. From the biological point of view, human cooperation is an evolutionary puzzle. Unlike other creatures, humans cooperate with genetically unrelated individuals, with people they will never meet again, when reputation gains are small or absent and even engage in altruistic punishment of defectors. These patterns of cooperation cannot be explained by kin selection, signaling theory or reciprocal altruism. The idea that group selection might explain this behavior goes back to Darwin himself who, in chapter 5 of the *"Descent of man and selection in relation to sex"*, states that "*... an increase in the number of well-endowed men and an advancement in the standard of morality will certainly give an immense advantage of one tribe over another.*" However, this idea felt in disrepute because evolution does not pitch groups again groups, nor individuals against individuals, but genes against genes. Then, a "selfish gene" analysis makes the altruistic good-of-the-group outcome virtually impossible to achieve. In particular because the late Pleistocene groups of modern man were not believed to be genetically sufficiently different to favor group selection. Therefore, human cooperation remained an evolutionary puzzle.

Contrary to the conventional self-interest behavior, postulated in economics (Homo Economicus), human cooperation in such extreme forms led Bowles and Gintis to develop the notion of *strong reciprocity* (*Homo Reciprocans* [4] [5]) as a better model for human behavior. *Homo Reciprocans would come to social situations with a propensity to cooperate and share but would respond to selfish behavior on the part of others by retaliating, even at a cost to himself and even when he could not expect any future personal gains from such actions*. This should be distinguished from cooperation in a repeated game or reciprocal altruism or other forms of mutually beneficial cooperation that can be accounted for in terms of self-interest.

Clear experimental evidence for the relevance of this notion is obtained from experimental games played by humans, which have outcomes very different from the self-interest (Nash equilibrium) result. A situation that is simple to analyze is the ultimatum game[2]. As a very interesting example, a group of anthropologists and economists conducted an "ultimatum game experiment" in many small-scale societies around the world [6] [7]. Nash equilibrium (Homo Economicus) was rejected in all cases, with different results obtained in different societies, the players' behavior being strongly correlated with existing social norms and the market structure in their societies. This strongly suggests [8] [9] that human decisions involve a mixture of self-interest and a background of (internalized) social norms [10] [11].

---

[2] The ultimatum game has two players, the *proposer (P)* and the *responder (R)*. The proposer is given an amount of goods which he can either divide into two equal parts (*b+b*) or into one large part for himself (*a*) and a very small amount (*c*) for the responder ($a + c = 2b, a \gg c$). If the responder accepts the division, each player receives his share. If the responder refuses the division, nothing is given to the players. The rational Nash equilibrium is that the responder should always accept, irrespective of the amount that is offered. However, when the game is played with human players, greedy proposals are most often refused, even in one-shot games where the responder has no material or strategic advantage in refusing the offer.

Strong reciprocity is a form of altruism [12] in that it benefits others at the expense of the individual that exhibits this trait. Monitoring and punishing selfish agents or norm violators is a costly (and dangerous) activity without immediate direct benefit to the agent that performs it. It would be much better to let others do it and reap the social benefits without the costs. Strong reciprocator agents contribute more to the group than selfish ones and they sustain the cost of monitoring and punishing free riders. For this reason, it was thought that the strong reciprocity trait could not invade a population of self-interested agents, nor could it be evolutionary stable. To check this hypothesis Bowles and Gintis [5] developed a simple model that might apply to the structure of the small hunter-gatherer bands of the late Pleistocene, which corresponds to about 95% of the evolutionary time of modern man. Taking the view that the *strong reciprocity* trait might have a genetic basis, this would be a period long enough to account for a significant development in the modern human gene distribution. The model would give an evolutionary explanation of the phenomenon. Of course, if instead of gene-based, strong reciprocity is culturally inherited, emergence and (or) modification of this trait could be much faster. Here I will analyze a simplified version of the Bowles and Gintis model[3], already used in Ref.[13] to study the evolution of strong reciprocity in large societies.

Let a population of size $N$ have two types of agents, one denoted *reciprocators* (R-agents) and the other *self-interested* (S-agents). In a *public goods* activity each agent can produce a maximum amount of goods $q$ at cost $b$ (with goods and costs in fitness units). The benefit that an S-agent takes from shirking public goods work is by avoiding the cost of effort $b(\sigma)$, $\sigma$ being the fraction of time the agent shirks. The following conditions hold

$$b(0) = b, \qquad b(1) = 0, \qquad b'(\sigma) < 0, \qquad b''(\sigma) > 0 \qquad (1)$$

Furthermore $q(1-\sigma) > b(\sigma)$ so that, at every level of effort, working helps the group more than it hurts the worker.

For $b(\sigma)$ choose

$$b(\sigma) = \frac{2}{2\sigma - 1 + \sqrt{1 + 4/b}} - \frac{2}{1 + \sqrt{1 + 4/b}} \qquad (2)$$

which satisfies the constraints (1).

R-agents never shirk and punish each free-rider at a cost $c\sigma$, the cost being shared by all R-agents. For an S-agent the estimated cost of being punished is $s\sigma$, punishment being ostracism or some other fitness decreasing measure. Punishment and cost of punishment are proportional to the shirking time $\sigma$ and $c$ is the reciprocator unit of punishment cost. $s$ is the weight given by an S-agent to the punishment probability. It may or may not be the same as the actual fitness

---

[3] The main simplification being that migration in and out of the evolving group, to an outside pool of agents, is not considered.

cost of punishment. Each S-agent chooses σ (the shirking time fraction) to minimize the function

$$B(\sigma) = b(\sigma) + sf\sigma - q(1-\sigma)\frac{1}{N} \quad (3)$$

f being the fraction of R-agents in the population, $f\sigma$ is the probability of being monitored and punished. The last term is the agent's share of his own production. The value $\sigma_S$ that minimizes $B(\sigma)$ is

$$\sigma_S = \max\left(\min\left(\frac{1}{2} - \sqrt{\frac{1}{4} + \frac{1}{b} + \frac{1}{\sqrt{sf + \frac{q}{N}}}}, 1\right), 0\right) \quad (4)$$

The contribution of each species to the population in the next time period is proportional to its fitness π given by

$$\begin{aligned}\pi'_S(f) &= q(1-(1-f)\sigma_S) - b(\sigma_S) - \gamma f \sigma_S \\ \pi'_R(f) &= q(1-(1-f)\sigma_S) - b - c(1-f)\frac{N\sigma_S}{Nf}\end{aligned} \quad (5)$$

for S- and R-agents. The baseline fitness is zero, that is,

$$\pi_{S,R} = \max\left(\pi'_{S,R}, 0\right)$$

The first term in both π'<sub>S</sub> and π'<sub>R</sub> is the benefit arising from the produced public goods and the second term the work effort. The last terms represent the fitness cost of punishment for S-agents and the cost incurred by R-agents. γ = 1 corresponds to ostracism from the group, other values to general coercive measures affecting the fitness of S-agents. The last term in π<sub>R</sub> emphasizes the collective nature of the punishment. Notice the heavy punishing burden put on reciprocators when in small numbers.

Finally, one obtains a replicator-dynamics one-dimensional map for the evolution of the fraction of R-agents

$$f_{new} = f\frac{\Pi_R(f)}{(1-f)\Pi_S + f\Pi_R(f)} \quad (6)$$

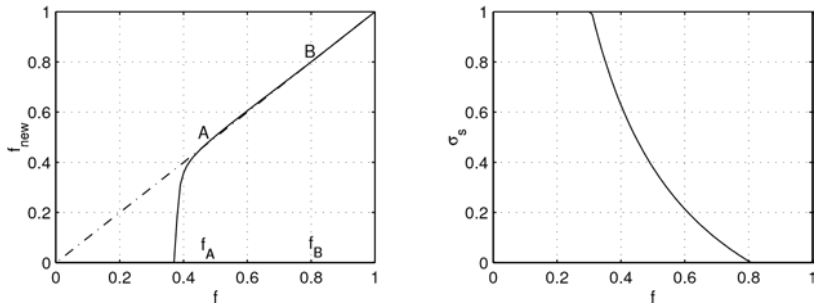

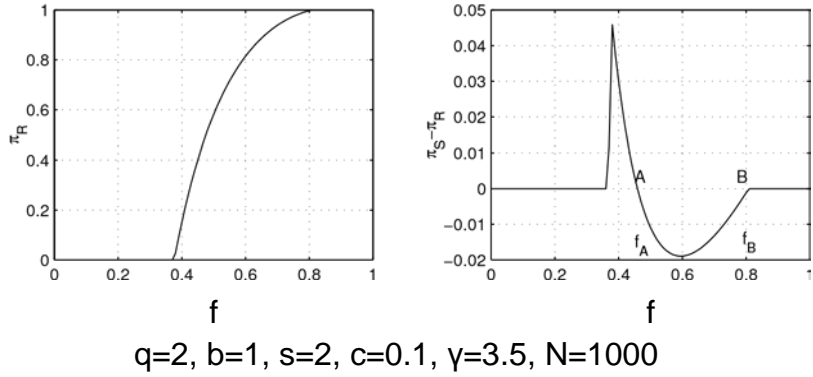

q=2, b=1, s=2, c=0.1, γ=3.5, N=1000

Figure 1: One-dimensional map for the evolution of R-agents, shirking times $\sigma_S$ and fitnesses $\Pi_R$ and $\Pi_S - \Pi_R$ as a function of the fraction $f$ of R-agents

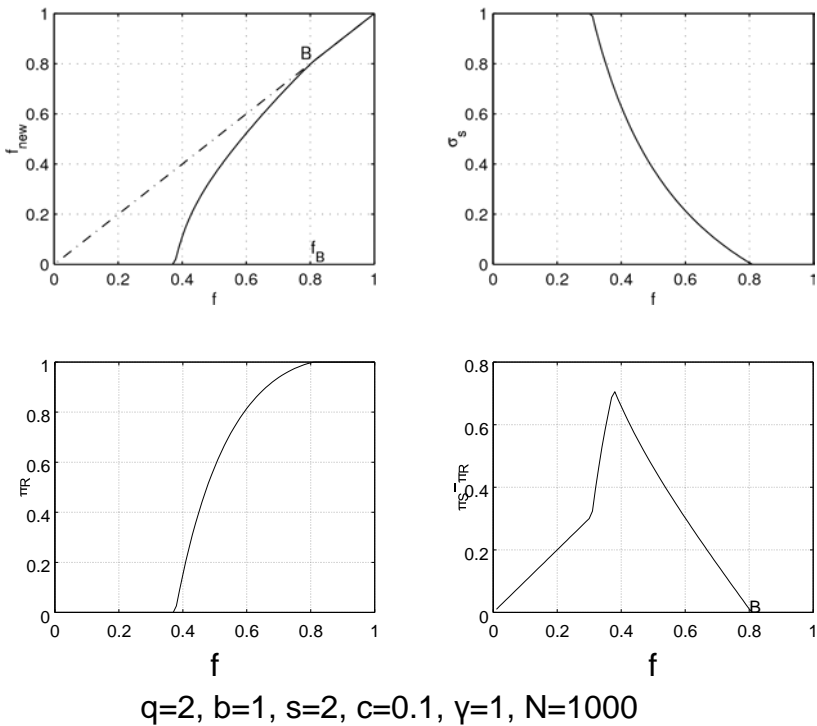

q=2, b=1, s=2, c=0.1, γ=1, N=1000

Figure 2: Same as Fig.1 with a different fitness cost of punishment

Figs.1 and 2 display this map, as well as $\sigma_S(f)$, $\Pi_R(f)$ and $\Pi_S(f) - \Pi_R(f)$ for two different values of $\gamma$, the other parameters being the same. They show the general behavior of the map in Eq.(6). If $\gamma$ (the fitness impact of punishment) is large enough, the map has an unstable fixed-point $A$ at $f_A$ and a left-stable one $B$ at $f_B$. Between $f_B$ and 1 there is a continuum of marginally stable fixed points. For smaller $\gamma$ the region between $f_A$ and $f_B$ (where $\Pi_S - \Pi_R$ is negative) disappears and only the marginally stable fixed points remain. In both cases the asymptotic behavior corresponds either to $f = 0$ (and $\sigma_S = 1$) or to $f$ between 0 and 1 but

$σ_S = 0$. That is, in this second case, both reciprocators and shirkers remain in the population, but shirkers choose not to shirk because the minimum of $B(σ)$ is at $σ_S = 0$.

For an initial $f$ smaller than $f_A$ the fraction of reciprocators falls very rapidly to zero. This reflects the fact that in this case a very small number of reciprocators must carry the burden of punishing very many shirkers. Hence, from the point of view of intragroup dynamics, either reciprocators are completely eliminated from the population, or they remain in equilibrium with a probably large number of shirkers, which do not shirk for fear of being punished. Notice also that the marginally stable nature of the (above $f_B$) fixed points implies that the shirker trait is never eliminated and will remain in the population.

**In conclusion:** *intragroup dynamics*, by itself, cannot explain how the reciprocator trait might have become dominant and cannot explain why humans became a cooperative species. It would seem that cooperation, either gene-coded or culturally inherited, is not evolutionary stable.

Furthermore, it is well known that group size affects monitoring in public goods provision [14]. Therefore, the situation would be even worse if, instead of the small hunter-gatherer groups, reciprocators and their fellow shirkers are members of a larger society. Monitoring and punishment of shirkers by reciprocators necessarily loses its global collective nature. Once monitoring loses its global nature, it might only become a chore for the neighbors of the shirker. In addition to the individual cost of monitoring and (or) punishing free riders, such punishing requires an amount of force that, in particular, ensures the effectiveness of the punishment and, on the other hand, puts the punisher safe from direct retaliation.

The effect of the social network structure on the evolution of the strong reciprocity trait was explored in Ref.[13] both as an agent-based and a mean field model. The results show the even greater fragility of cooperation in intra-group dynamics. In the new model the monitoring function performed by R-agents operates at the neighbors' level and punishment is only implemented if at least two neighbors are willing to do so. It is the same as to say that punishing a norm-violator cannot be an individual action but requires a minimum of social power and consensus. The need to be close by and the need for agreement of at least two neighboring reciprocators to implement punishment, suggests that the structure of the social network plays a role on the evolution of the group. The mathematical coding of this idea is as follows:

As before, there are two agent types (S-agents and R-agents), the fraction of R-agents being $f$. The agents are placed in a network where, on average, each agent is connected to $k$ other agents. $k$ is called the *degree* of the network. To the whole population of dimension $N$, one associates 3 $N$−dimensional vectors, *Wk*, *Pu* and *Cpu*. *Wk* is called the *work vector*, *Pu* the *punishment vector* and *Cpu* the *cost of punishment vector*. The link structure of the network is chosen as in the *β*−model of Watts and Strogatz[15].

Each reciprocator, on detecting an S-agent $k$, looks for another reciprocator in his own neighborhood also connected to $k$. If he finds one, he punishes $k$ by an amount proportional to the fraction of shirking. An S-agent may be punished several times by all different pairs of reciprocators in his neighborhood. The amount of work that an S-agent does is inversely proportional to the number of reciprocators in his neighborhood. However, lack of communication between neighboring reciprocators may make the probability of punishment much smaller.

The (average) fitness of R-agents and S-agents is

$$\pi'_R = \frac{q}{N}\sum_{all} Wk(i) - \frac{b}{fN}\sum_{i \in R} Wk(i) - \frac{c}{fN}\sum_{i \in R} Cpu(i) \quad (7)$$

$$\pi'_S = \frac{q}{N}\sum_{all} Wk(i) - \frac{b}{(1-f)N}\sum_{i \in S} Wk(i) - \frac{\gamma}{(1-f)N}\sum_{i \in S} Pu(i) \quad (8)$$

The baseline fitness is zero, that is

$$\pi_{R,S} = \max\left(\pi'_{R,S}, 0\right) \quad (9)$$

with the constants $q$, $b$, $c$ and $\gamma$ having the same meaning as in the model studied before. Namely, $q$ is the maximum amount of goods that each agent can produce at cost $b$, $c$ is the cost of punishment and $\gamma$ the fitness cost of being punished. Notice however that $c$ is now the cost of punishing to one punisher, not the cost for each pair of punishers ($2c$). The total social cost might even be larger if the S-agent has more than one pair of R-agents in his neighborhood.

Once the fitness is computed the replicator equation

$$f_{new} = f\frac{\pi_R}{f\pi_R + (1-f)\pi_S} \quad (10)$$

is applied and a new cycle starts with a new random distribution, on the network, of $Nf_{new}$ R-agents and $N(1 - f_{new})$ S-agents.

Running this agent model for several values of $\beta$ and, in each case, for random initial $f_0$'s one finds two separate regions in the ($f_0,\beta$) plane (Fig.3). In region 1 the evolution drives $f$ towards zero as well as the overall fitness $\pi$ (as shown in Fig.4a)

$$\pi = f\pi_R + (1 - f)\pi_S \quad (11)$$

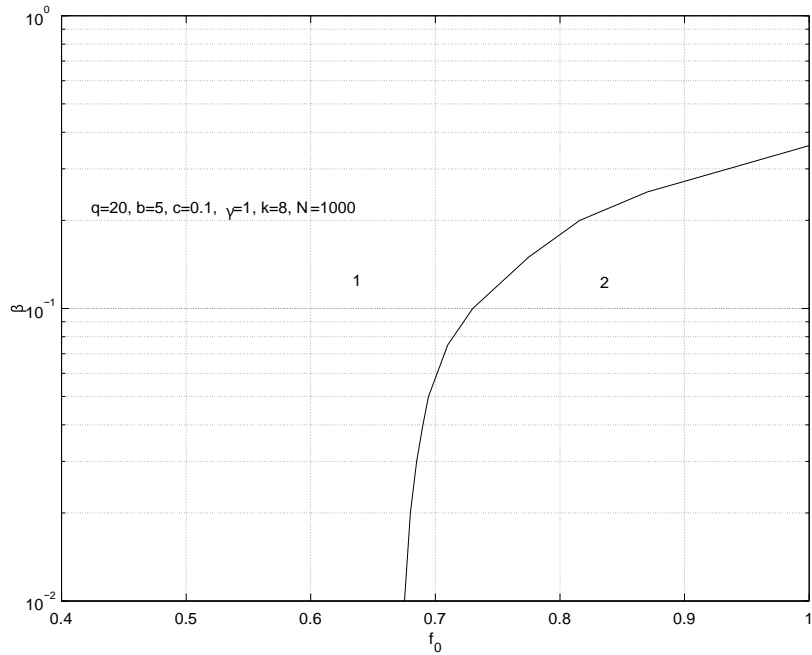

Figure 3: The two-phase regions in the ($f_0$, $\beta$) plane

In region 2 there is an asymptotic nonzero value for $f$ and for the fitness (shown in Fig.4b).

As $\beta$ increases it becomes less likely to have a stable nonzero $f$. The origin of this effect is clear. Although $\beta$−rewiring maintains the average degree of the network, the probability of two neighbors of an agent to be themselves neighbors decreases. Therefore, it becomes increasingly difficult for reciprocators to find local consensus for the punishment of S-agents.

The average probability of two R-neighbors of a S-agent to be themselves neighbors, is called the (relative) *clustering coefficient*, the network clustering coefficient being related to the notion of transitivity used in the sociological literature. For the $\beta$−rewiring model, the clustering may be estimated from the number of shortcuts, which in this case is proportional to $\beta$ [15]. This allows us to establish a mean field version of the agent model (see Ref.[13] for details).

The mean field version gives results identical to the agent-based model,

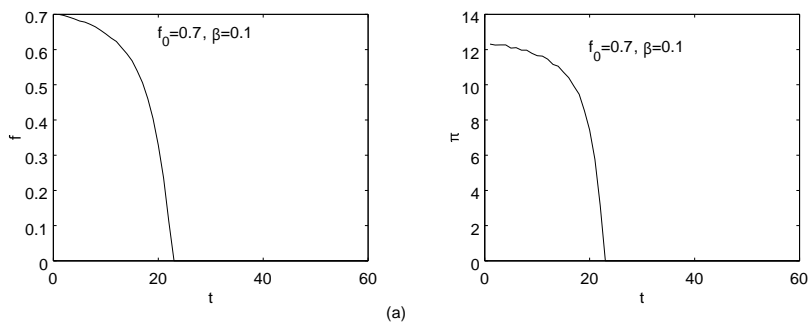

(a)

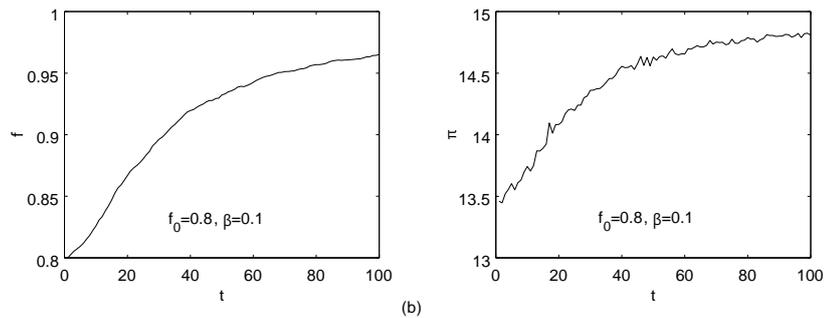

Figure 4: Evolution of the fraction *f* of R-agents and global fitness in region 1 (a) and region 2 (b)

clustering appearing as the determining network parameter that drives the evolution of the reciprocator trait. The general conclusion is that *intragroup dynamics cannot explain the evolutionary origin of the strong reciprocator trait* and whatever the origin of this trait was, one sees that this type of cooperative behavior is even more difficult to maintain in larger societies.

## II B. Conflict and parochial cooperation (P_cooperation)

Intragroup dynamics cannot explain the evolutionary emergence of the type of cooperative behavior displayed by the human species. Nevertheless, it exists, and some evolutionary phenomenon must have led to it. Two hypotheses were put forward to explain the paradox:

1 - When very many groups are considered, for example assembled at random from a pool containing both reciprocators and shirkers [16] [17], then only the groups that contain at the start a fraction *f* greater than $f_A$ will have in the end a nonzero fitness. In all others, S-agents invade the population and suffer a *tragedy of the commons* situation, with final zero fitness. Hence, groups with reciprocators might still tend to dominate and impose an above average predominance of the reciprocator trait.

2 - A more realistic hypothesis was put forward by Choi and Bowles[18]. They consider a particular type of cooperation between humans, where *they are able to benefit fellow group members, even at a cost to themselves, whereas hostility is frequent towards individuals not of the same ethnic, racial or other group*. This is what they call *parochial altruism*. From their analysis it follows that neither parochialism nor altruism would have been evolutionary viable by themselves but, by promoting conflict between different groups, they could have evolved jointly. Groups where internal cooperation would have developed would have a clear advantage in conflict situations. In conclusion: *intragroup cooperation and war are born at the same time* [19] [20].

*Parochial cooperation*, that is, the tendency to cooperate and extend trust towards those that belong to one's own group (in-group love) and a willingness to fight against rivaling groups (out-group hate), has been extensively studied [21]

[22] [23] [24] [25] [26] [27]. As one these authors points out, the selfless sacrifice of a man to save Jewish children from the holocaust and the murderous blow-up of a Chechen woman in a Russian public place, presumably to hurt Russians for cruelty and oppression, are in fact two faces of the same coin. There are also biological and experimental hints [24] [27] that this human trait may already be genetically encoded and, therefore, quite resistant to cultural intervention.

The human ability to work together surpasses that of any other species and cooperation seems to be the foundation of our progress. The history of humanity is one of battles, violence and genocide, a trend which at first sight would seem to contradict the survival instinct. We entertain ourselves with sports in which victory is marked by humiliation of others, create economic systems based on inequalities of power, all the time surrounded by simulations of war. The great engine of human sociality, paradoxically, was conflict itself. The realization that this is a dominant trend in human history is very old. Already Heraclitus in one of his surviving fragments says "*War is the father of all things, king of all things and some he revealed as gods, others as men, he made some slaves, others free*". And war itself and its heroes have at times been so glorified[4].

Harari [28] asserts that "imagined orders", that is, the set of imagined realities, common stories and shared myths that humans create to organize society, is what enables humans to cooperate with strangers on a large scale. This set of human-created myths is what men invented to extend parochial cooperation to larger groups. And because parochial cooperation seems to be biologically encoded [24] [27], imagined orders could, after all, also have a biological origin. The danger, of course, is that imagined orders, rather than being created by gods or laws of Nature, are human creations and, therefore, there are large variety of them, creating conflicting tribes.

Parochial cooperation (here denoted **P_cooperation**) and **Conflict** are the first two pillars of the behavior pattern of Homo Sapiens. They have, along history provided the most performing groups with success and growth, growth in territory and wealth. And growth itself has become a paradigm of all aspects of human life. So **P_cooperation, Conflict and Growth** became the dominant behavior pattern in the Human Planet.

## II C. Emergence of government and the "tragedy of authorities"

Frequent warfare makes altruistic cooperation among group members essential to survival. However, maintaining cooperation in peaceful times requires monitoring of free riders which, as seen before, are never eliminated from the population. As the groups become larger, the size and degree of clustering precludes the collective nature of rule-violators monitoring. Emergence of group-level social norms would be beneficial, at least for R-agents. But who is going to control them? The transition from the small hunter-gatherer groups to larger sedentary population groups occurred at the time of the agricultural revolution and the solution was *the emergence of government*[5]. That is, a new type of agent (*the ruler, the authority*) came into play and replaced the type of egalitarian decision-making that might have existed before. It is worth noticing that the apparent ease with which humans accepted this transition of power may have to do with the internalization of the reciprocator trait that, above complete freedom, valued the enforcing of social norms.

---

[4] It is a pure sentimental illusion to expect too much from humanity if it forgets the art of war. Nothing but war can awaken in an exhausted people, in a strong and sure way, that battlefield energy, that deep and impersonal hatred, that cold-blooded murderer of good conscience, that communal and organized ardor to destroy the enemy, that proud indifference to the great losses of one's own existence or that of friends, that silent convulsion, like an earthquake of the soul. Everything a people needs when it is losing its vitality (Nietzsche, *Human all too Human*)

In the agricultural societies, specialization arose as well as new security needs and there was a more intense population pressure on limited resources. This tended to create greater organization within the community, which in turn led to social hierarchies, to certain forms of chieftainship and to a whole class of people with managing roles. The government rulers that the first agricultural societies accepted might have been priestly figures, which might have depicted themselves as servants of the gods, acting on behalf of the community. The emergence of organized religion appears as a norm-enforcing tool, because it is easier for the ruler to invoke the will of the gods than their own personal preferences. As time went by, especially because of conflict with other rival city states, the cities came to rely more and more on military leaders and the ruling priests gave place to military leaders.

As before, the dynamics of this new situation may be analyzed by a simple model [29]. Authority agents appear to avoid a "tragedy of the commons", that is, a fitness crisis arising from the proliferation of free-riding agents. The interesting question is that the dynamics of the authority agents may, by itself, lead to a new fitness crisis which has been called a *tragedy of authorities*. The basic setting is similar to the one used before [13] with a *public goods activity* in a group of $N$ agents, $N$ being in general a function of time. Three types of agents are considered. The first type (*R-agents*) are cooperators that also have a monitoring effect on the cooperation of other agents. The second are self-regarding agents (*S-agents*) and the third are purely monitoring agents (*A-agents*). The labels that are chosen refer to the names reciprocators (R), self-regarding or shirkers (S) and authorities (A). The percentages of each one of the types in the population are denoted by $f_R$, $f_S$ and $f_A$.

Each R or S-agent can produce a maximum amount of goods $q$ at cost $b$ (with goods and costs in fitness units). An S-agent benefits from shirking public goods work by decreasing the cost of effort $b(\sigma)$, $\sigma$ being the fraction of time the agent shirks. The conditions on $b(0)$, $b(1)$, $b(\sigma)$, $b'(\sigma)$ and $b''(\sigma)$ are the same as in Eqs. (1) and (2).

---

[5] In the models, the fact that the population fraction of reciprocators is not invasive and the impossibility of collective monitoring in large populations, makes *emergence of government* a mathematical necessity.

R-agents never shirk and punish each free-rider at cost $c\sigma$ with probability $p(N)$, the cost being shared by all R-agents. For an S-agent the estimated cost of being punished is $s\sigma$. Punishment and cost of punishment are proportional to the shirking time $\sigma$ and $c$ is the reciprocator unit of punishment cost. $s$ is the weight given by an S-agent to the possibility of being punished. It may or may not be the same as the actual fitness costs of punishment ($\gamma, \gamma_A$). Each S-agent chooses $\sigma$ (the shirking time fraction) to minimize the function

$$B(\sigma) = b(\sigma) + s(f_R + f_A)\sigma - q(1-\sigma)\frac{1}{N} \qquad (12)$$

From the point of view of an S-agent $(f_R + f_A)\sigma$ is the probability of being monitored and punished. The last term is the agent's share of his own production. The value $\sigma_S$ that minimizes $B(\sigma)$ is

$$\sigma_S = \max\left(\min\left(\frac{1}{2} - \sqrt{\frac{1}{4} + \frac{1}{b}} + \frac{1}{\sqrt{s(f_R + f_A) + \frac{q}{N}}}, 1\right), 0\right) \qquad (13)$$

The contribution of each species to the population in the next time period is proportional to its fitness $\pi_R$, $\pi_S$ or $\pi_A$ computed from

$$\begin{aligned}
\pi'_R &= q(1 - f_A - f_S\sigma_S)x - b - cp(N)f_S\frac{N\sigma_S}{Nf_R} \\
\pi'_S &= q(1 - f_A - f_S\sigma_S)x - b(\sigma_S) - (\gamma p(N)f_R + \gamma_A f_A)\sigma_S \\
\pi'_A &= q(1 - f_A - f_S\sigma_S)wx - c_A f_S\frac{N\sigma_S}{Nf_A}
\end{aligned} \qquad (14)$$

with

$$\pi_{R,S,A} = \max\left(\pi'_{R,S,A}, 0\right)$$

because the baseline fitness is zero.

The first term in both $\pi'_R$, $\pi'_S$ and $\pi'_A$ is the benefit arising from the produced public goods. The factors $x$ and $wx$ with

$$x = \frac{1}{wf_A + 1 - f_A}$$

account for the fact that this benefit is the same for R and S-agents but might be different for A-agents. The second term in $\pi'_R$ and $\pi'_S$ is the work effort. The third term in $\pi'_R$ and the second term in $\pi'_A$ represent the fitness cost of punishment for R and A-agents and the third term in $\pi'_S$ the cost incurred by S-agents when they are punished. The $\gamma$ and $\gamma_A$ coefficients code for the severity of the coercive measures affecting the fitness of S-agents. The last term in $\pi'_R$ and $\pi'_A$ emphasizes the heavy punishing burden put on R or A-agents when in small number. The factor $p(N)$, a decreasing function of $N$, accounts for the fact that (as studied at length in [13]), when a social group grows in size, the collective nature of monitoring of free-riders becomes increasingly difficult. Essentially, the punishment probability by R-agents should be a growing function of the clustering

coefficient of the group. Here, for illustration purposes, a simple function of $N$ is chosen

$$p(N) = \sqrt{\frac{1+\delta}{1+\delta\frac{N}{N_0}}}$$

$N_0$ being some small initial population.

Finally, the evolution of the population at successive generations, is obtained from the replicator map[6]

$$f_{\alpha,new} = f_\alpha \frac{\Pi_\alpha(f)}{f_R \Pi_S + f_s \Pi_S + f_A \Pi_A} \qquad (15)$$

$\alpha = R, S, A$.

When $f_A = 0$ and $p(N) = 1$ the map has an unstable fixed point at $A$ ($f_R(A) \simeq 0.57$), a left-stable fixed point at $B$ ($f_R(B) \simeq 0.85$) and a continuum of neutral fixed points after that. For $p(N) = 0.5$ only the neutral fixed points remain (Fig.5). The neutral fixed points correspond to the situation where S-agents do not shirk for fear of being punished. For initial conditions smaller than $f_R(A)$ in the first case or $f_R(B)$ in the second, the population of R-agents is always invaded by S-agents. However the neutrality of the fixed points means that the population of S-agents is not completely invaded by the R-agents.

Next, still keeping $f_A = 0$, the evolution of the population of R and S-agents was studied when the population increases in time according to a global fitness dependent law, chosen as

$$N(t+1) = N(t)e^{\beta\pi}$$

with $\pi = \Sigma_\alpha f_\alpha \pi_\alpha$.

Fig.6 displays the results for a time-evolution starting from $N_0 = 20$, $f_R = 0.7$, $f_S = 0.3$. In the upper left plot, the percentages $f_R, f_S$ and $f_A$ ($f_A = 0$ in this case) of each agent type are displayed as the distances to the three sides of a triangle. One sees that as long as the population ($N$) remains small the monitoring effects of R-agents controls shirking ($\sigma$) by the S-agents and, as a result, their percentage ($f_R$) and fitness ($\Pi_R$) increases as well as the average fitness of the group. However, with further population growth the punishment probability ($p(N)$) of shirkers decreases leading for a while to a higher degree of shirking ($\sigma$) and higher fitness ($\Pi_S$) and percentage ($f_S$) of S-agents. But because S-agents with high $\sigma$ produce much less goods, finally the fitness of all agents decreases and the group collapses. This is the well-known *tragedy of the commons*.

---

[6] A different, incremental, dynamics is sometimes used for the fitness-based evolution of populations. The replicator map used here provides faster evolution but qualitatively similar results, up to a renormalization of the time scale.

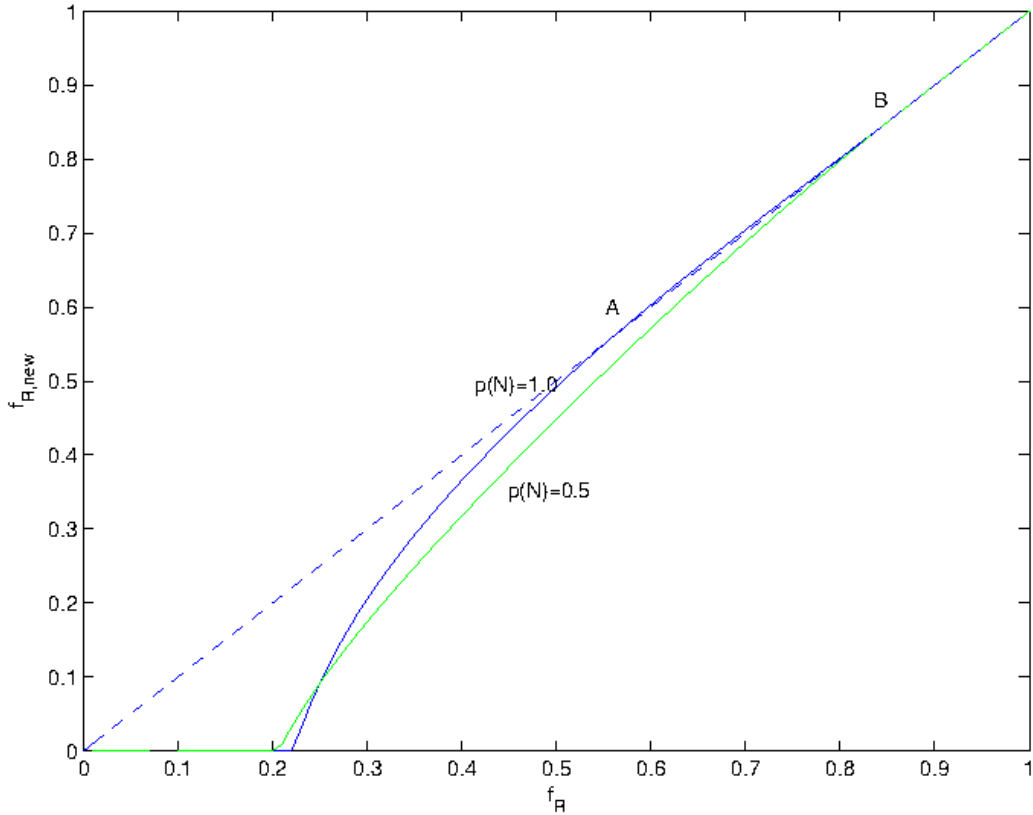

Figure 5: One-dimensional map for the evolution of R-agents corresponding to $f_A = 0, q = 2, b = 1, c = 0.1, y = 4, s = 3, N = 20$

It is then natural that a population group whose success is based on cooperation and control of selfish behavior, would recognize the need, beyond a certain population level, to assign the control and punishing role to specialized agents, with extra power and authority. This is what has been called the *emergence of government*. In the model, one now starts from the same initial conditions, but when $f_R$ reaches a value below 0.5 unfreeze the dynamics of A-agents, imposing however, for the moment, the constraint that $f_A$ should not exceed 0.2 and, to isolate the effect of the A-agents, the population is assumed to be constant after that moment. The result is shown in Fig.7.

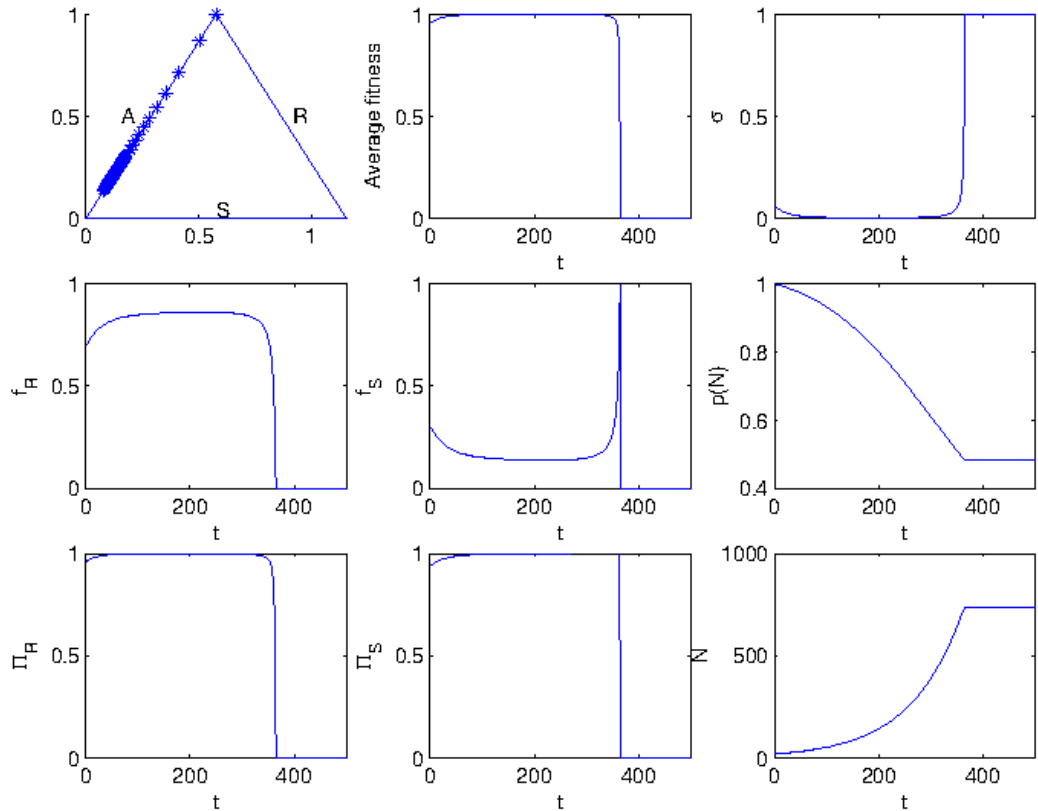

Figure 6: Time evolution of R- and S-agents with $f_A = 0$, $q = 2$, $b = 1$, $c = 0.1$, $\gamma = 4$, $s = 3$, $N_0 = 20$

The outcome is satisfactory. After the unfreezing of the $f_A$ dynamics the percentage of R-agents still decreases for a while, but then it stars to grow and the group stabilizes at an high level of average fitness.

Notice that the growth of the number of A-agents is rather fast. The reason is that as soon as they start controlling the behavior of the S-agents, both $\sigma$ and $f_S$ decrease, therefore greatly increasing the fitness of the agents, because they benefit from the goods produced without incurring the cost of control because there is almost nothing to control anymore. If one now removes the 0.2 bound on $f_A$ (Fig.8) the A-agents population continues to grow but, because they produce no goods, the average fitness finally decreases to zero as the group collapses. This is a crisis of a different type, that one might call the *tragedy of authorities*.

A very similar effect is obtained if, while keeping $f_A$ bounded, one allows $w$ to grow with the fitness of A-agents. That is, allowing the share of goods allotted to A-agents to grow.

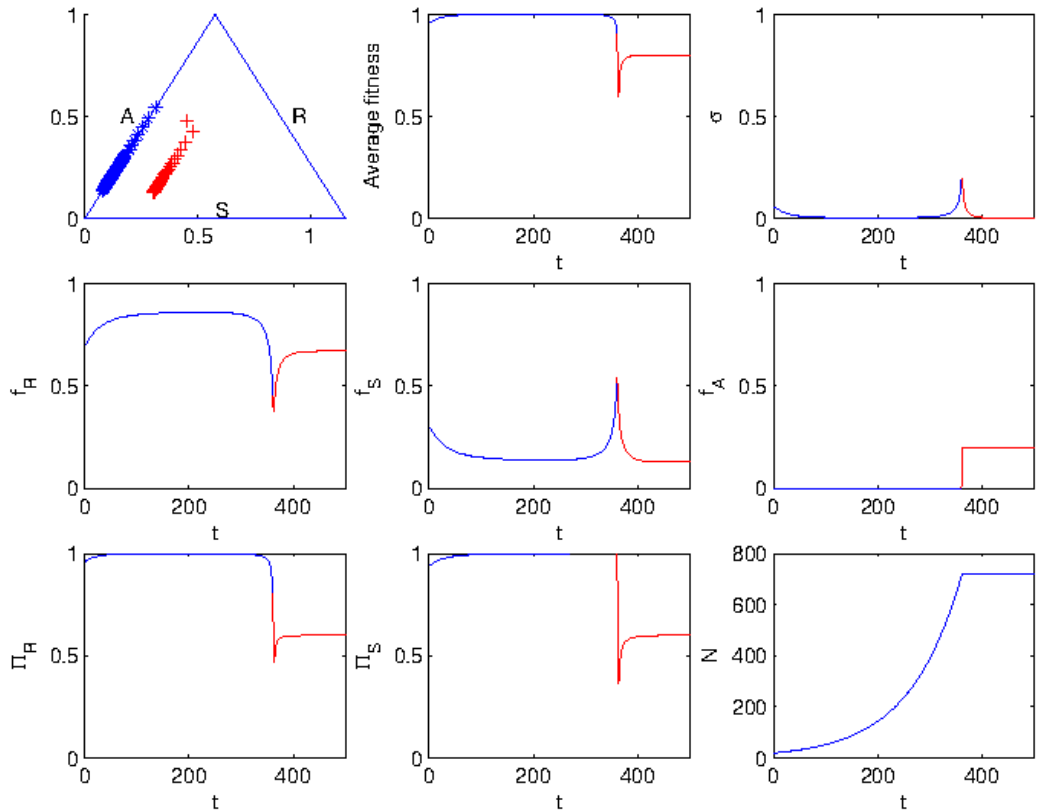

Figure 7: Time evolution with the three types of agents but $f_A \leq 0.2$.
($q = 2$, $b = 1$, $c = 0.1$, $s = 3$, $c_A = 0.45$, $γ = 4$, $γ_A = 11$)

This study emphasizes the delicate nature of the balance between the several agents in a viable society and the emergence of what seem to be universal features in the human social evolution. Cooperation is at the root of success in human groups. However, a natural, perhaps biological, tendency of humans to minimize effort and to maximize benefits requires that a certain amount of control of shirking is required. This led some humans to internalize the idea that shirkers should be controlled. Apparently, it is the societies where more humans adopted this norm that were the most successful. When, after the agricultural revolution the human groups became larger, collective control became more difficult. Then, the evolved acceptance of social norms led naturally to the acceptance of government as a specialized body. However, the dynamics of the authority agents may, by itself, lead to a new fitness crisis. The agents associated to governance, that is, the ruling class, may by its proliferation or by assigning to itself a higher

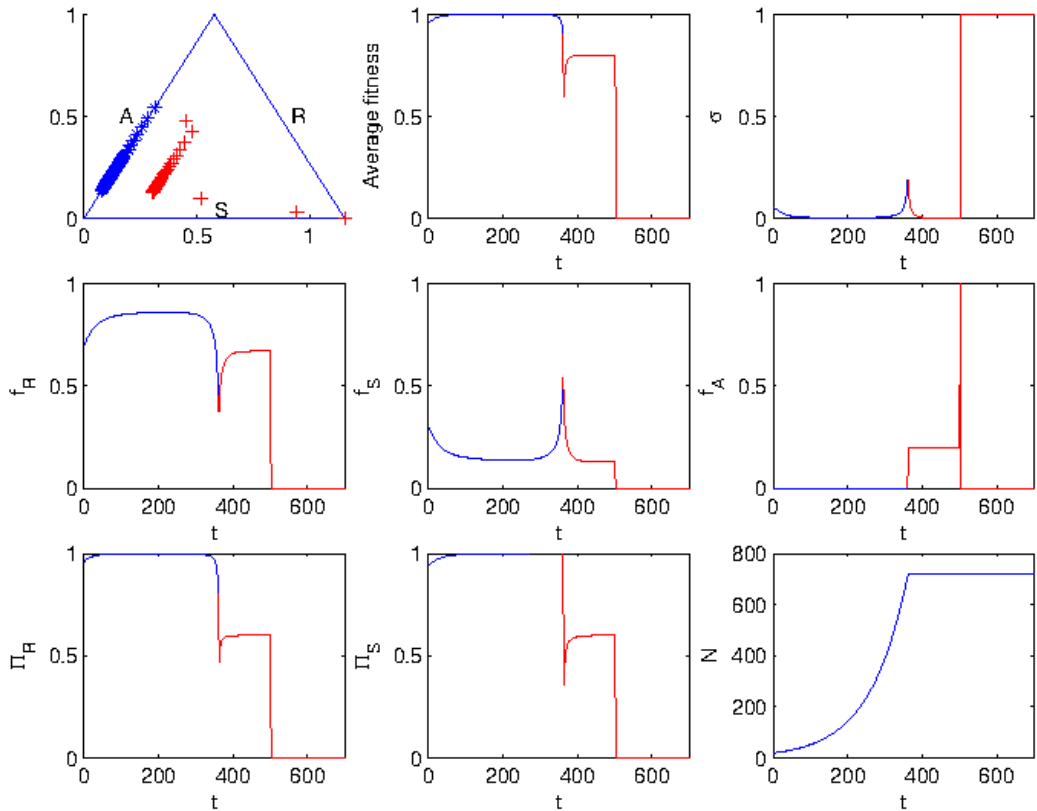

Figure 8: Time evolution with the three types of agents and $f_A$ allowed to grow above 0.2 after time 500. ($q = 2$, $b = 1$, $c = 0.1$, $s = 3$, $c_A = 0.45$, $\gamma = 4$, $\gamma_A = 11$)

share of the production (a high *w* factor in the model) provoke a decrease of the average fitness, a crisis or even a collapse of the society. This is what one calls the *tragedy of authorities*. It is a situation where authorities are not playing the role for which they have been created.

Some authors [30] [31] [32] have studied the historical effects of "elite overproduction" as generating crisis and revolutions. However not all cases of elite overproduction, that they characterize, can be identified with the phenomena of the tragedy of authorities. If elite overproduction is, for example, the proliferation of an aristocratic class, that under the protection of the ruler lives from the society production without contributing to it, then it has all the marks of a tragedy of authorities. But if, instead, elite overproduction is associated to an higher access of the youth to higher education, this is not a tragedy of authorities. The eventual crisis that may occur in this case results from the fact that the new educated agents are not incorporated neither in the productive sector nor as beneficiaries of the society production. Hence it is not a tragedy of authorities. In fact they are only reacting against an authority structure that wants to preserve their privileges. Therefore, to associate this two distinct situations under the same elite overproduction label may be quite misleading. The existence of authority agents is beneficial to society as long as their number and their share of the goods remains limited. The problem therefore is the old question of *who controls*

*the controllers*[7]. Democracy is in principle a way to implement limitations and accountability of the rulers. But even then, nothing is guaranteed. Economic power easily escapes the constraints of democratic control. And even more subtle effects may occur. For example, through exploration of the co-evolved parochial feelings of the population, it is easy to erect as a goal the proliferation of local or regional government structures, coordinating committees, etc. Layers and layers of control when there is nothing else to control. Autocratic forms of government are, of course, extreme cases of a tragedy of authorities, but even subtler effects are found everywhere. The solidary form of collective government of the hunter-gatherer groups was probably the most successful invention of modern man, leading to his dominance over other species and even over other hominids. It was also the most extensively tested of all, lasting for 95% of the evolutionary history of modern man. Centralized, professional forms of government, by comparison, are a very recent development, not always very successful. Hence, it could be rationally expected that, whenever applicable, "community government" would be used. In fact, and except in very rare cases this is not so. Instead, centralized forms of government tend to migrate to all local levels carrying with them the kind of political party-oriented issues, which are not necessarily the most relevant at the local community level.

Tragedy of authorities, in its more basic form of authoritarianism, has throughout history occurred very frequently [33]. Sometimes it is even consciously or unconsciously accepted or desired by the suffering communities [34]. At the origin, the purpose of the authorities is to insure, through the rule of law, the degree of cooperation that is needed for the collective good. Particularly interesting are the mechanisms that authorities, that are in an abusive tragedy of authorities situation, use to quiet their subjects when they are on the verge of rebellion. Of course, violent repression is the obvious answer. However, very often they resort to subtler devices. Devices that are exactly grounded on the knowledge of the behavior pattern, *P-Cooperation, Conflict, Growth*. *P-cooperation* has its maximal expression in the love of *nationality* and the hate of everybody else. They then play this cord, invent a symbolic external enemy or an elaborate foreign inspired conspiracy. Or because nationality and war are founding principles, they celebrate past military glories to mobilize the attention and divert their communities from the real problems. A collection of very interesting examples of this phenomenon is provided in a recent book by Katie Stallard [35]. Others [36] use the same social media manipulations that the so-called "spin doctors" [37] use to win improbable elections.

## III. The crises that are born from the behavior pattern

1. A first and persistent crisis, arising from the behavior pattern of HomoSapiens, is the *tragedy of authorities* discussed before. In its diverse forms of autocracy, exploitation and inefficiency, it stands on the way of a reasonable appropriation by the society of its material and immaterial goods. In addition, it is a crisis that is always waiting to happen, as is clear from the unstable nature of the 3-agents model discussed before. Referring to the three pillars of the behavioral pattern, this crisis is mostly related to the first one, the *P-Cooperation*.

---

7 *Quis Custodiet Ipsos Custodes* (Juvenal), although Juvenal himself has used it in a different context

2. *Conflict* is the second pillar and, throughout history, with violence, cruelty and genocide, it seems to have played a role as stabilizer of cooperation. As far as survival of a human group is concerned, the seriousness of a conflict is mostly related to the destructive nature of the instruments of war that are used. It so happens that, contrary to the past, the instruments of war that now exist are capable of extermination of the belligerent groups or even a global one. With the proliferation of nuclear capabilities in many places and the new flexibility of these weapons [38] [39] [40], a global nuclear conflict is a crisis that is becoming less and less improbable.

3. Finally, a new type of crisis arises from the third pillar, *Growth*. In the past, the human interventions in the planet had, either a local impact, or an impact much smaller than the regeneration capacity and the material resources of the planet. Not anymore. On a finite environment, any growing system, sooner or later, hits the boundaries.

Population growth on a finite planet cannot continue indefinitely for any living species. Living systems require energy for their existence and the proportion of available energy that is required for maintenance of the living systems increases as the size and complexity of those systems increase. For social species, increasing numbers require additional energy for maintaining their cohesion and livelihood. The resources that humans need to support food and shelter requirements are partly non-renewable. They require the extraction of various minerals, water, fuels, and building materials. Over time these resources decline and become increasingly difficult to extract. In the case of fresh water, supplies are becoming increasingly polluted. This not only affects humans directly, but also all the other species that constitute the basis for our food supplies. The availability of medicinal drugs and building materials is also affected. This problem is highlighted each year on Earth Overshoot Day [41], an illustrative calendar date obtained through calculation, on which day humanity's resource consumption for the year is considered to have exceeded the Earth's capacity to regenerate those resources for that year. In 2024, it fell on August 1st. For five months we will be deficit spending these resources, and in the process doing damage to the Earth's capability of generating these renewables.

Unlimited growth is also influencing weather patterns and the impact on other species is also very serious because human existence cannot succeed by going it alone. The presence of a rich biota is needed to provide the conditions for our survival [42].

In conclusion: Homo sapiens that in the past, in several places, has suffered natural ecological disasters [43], is now, by himself, engineering a global one.

## IV. Any hope?

Avoidance of the crises is, in a sense, synonymous to the fulfilment of the 17 Sustainability Goals. Most of the potential crises and current issues are global problems and, therefore, require global solutions. The first utopic thought that comes to one's mind is that **a new cultural paradigm is needed,** that is, cooperation within a single group, that group being the whole human planet. Is that possible?

There have been, in the past, proposals to establish such a paradigm. One would be **control by a central government?** However, there was no success of this idea. Let us look at the particular example of *conflict regulation*. Much before the League of Nations or the UN there were proposals for conflict regulation. For example, after the French-Polish war, in 1795, Kant proposed a plan for perpetual peace (*Zum ewigen Frieden. Ein philosophischer Entwurf*). Kant's proposals, the League of Nations Covenant or the United Nations Charter are all very similar. For example, the *Article 1 of the UN Charter* says:

- *To maintain international peace and security, and to that end to take effective collective measures for the prevention and removal of threats to the peace, and for the suppression of acts of aggression or other breaches of the peace, and to bring about by peaceful means, and in conformity with the principles of justice and international law, adjustment or settlement of international disputes or situations which might lead to a breach of the peace*;
- *To develop friendly relations among nations based on respect for the principle of equal rights and self-determination of peoples, and to take other appropriate measures to strengthen universal peace*;
- *To achieve international cooperation in solving international problems of an economic, social, cultural, or humanitarian character, and in promoting and encouraging respect for human rights and for fundamental freedoms for all without distinction as to race, sex, language, or religion; and*
- *To be a center for harmonizing the actions of nations in the attainment of these common ends.*

It suffices to look at the history of the second half of the 20th century and at the current situation of the world, to see that none of the good intentions and proposed aims have been achieved. The only few successes of the UN are actions of its specialized agencies, for very specific problems that do not confront the will of the nations.

The inefficiency of central control is quite evident. It is easy to understand why: Attempting to regulate the relations between nations, both the League of Nations and the United Nations were bound to fail. The behavior paradigm (**P-Cooperation-Conflict-Growth**) implies as a first attribute the strong idea of *national sovereignty*. The mimicry of "A Parliament of Men" that the UN is, collides at the outset with that strongest foundational idea. So, it is an *error of casting*. And mimicry it also is, based on the idea that all nations, independent of their size or world impact have the same one vote. A vote that, in any case, is irrelevant if it collides with the wishes of the security council.

Of course, the UN may be a useful discussion forum, to raise problems and bring complaints. But little more. The establishment of a new cultural paradigm, one of global world cooperation, collides with a behavior pattern that evolved over more than 70000 years. And there is some evidence that at least part of the behavior pattern is neurologically coded. May the urgency of current problems wait that long to uncode it? Hence, any hope for a solution should take into account the current behavior pattern and not, at the outset, collide with it. An interesting possibility, not colliding with the P-cooperation trait and its strong corollary, the national identity, has been described in a recent book by Mangabeira Unger [44]. Instead of a parliament of men, the chance for success might be the formation of "Coalitions of the Willing" to independently address each one of the goals. Independent coalitions, with independent agencies operating only in the framework of each one of the particular SDG. Even partly successful existing agencies, like the International Court of Justice or the WTO should be independent agencies, not operating under the aegis of the UN. *Governance without government*, to establish *a network of voluntary complicities* between the nations, rather than central control by a pseudo-world government.

The other pillars of the behavior pattern also deserve attention. *Growth* has been a paradigm of success for all kinds of human groups. And, although some less-evolved individuals still dream of territorial growth for their tribe, the great majority aims for economic growth. And economic growth very often disregards any concerns over sustainability or conflict with the boundaries of Nature. Gross Domestic Product (GDP), the sum of consumption, investment, government spending and net exports, is the most popular indicator of a nation's economic performance. To make GDP grow is a sacred vow of all rulers. But even for the economy, GDP is an imperfect measure of overall economic well-being [45]. Alternative metrics such as the Human Development Index, BetterLife Index or the Genuine Progress Indicator have been proposed. And although keeping intact the growth pattern trait, some of these metrics, if adopted, might be less deleterious in what concerns the production of pollution and the overexploitation of the planet resources while, at the same time, leading to better indicators for the health and education of the populations.

Finally, *Conflict* is unavoidable, when resources are limited and the ambition of its appropriation is unlimited. Therefore, all we might aim for, is the minimization of the conflict consequences. To regulate conflict, to reach a middle ground agreement, without resorting to armed conflict or other extreme measures. Conflict and conflict regulation is a science by itself [46]. An international, non-aligned, agency of experts in this science, offering its advising services to the conflicting parts might be a better choice than the security council of the UN, always hostage of the delicate balance of the powers. Once a truce is reached, maintenance of the peace, as implemented by the UN, is also poorly designed. A peacekeeping force composed of troops that were trained for war and not to keep the peace is a contradiction. A special, permanent body of peacekeeping agents, specially equipped and trained for nonlethal intervention, would be a better choice.

### Acknowledgments

Partially supported by Fundação para a Ciência e a Tecnologia (FCT), project UIDB/04561/2020: https://doi.org/10.54499/UIDB/04561/2020